\documentclass[conference]{IEEEtran}
\IEEEoverridecommandlockouts
\usepackage{cite}
\usepackage{amsmath,amssymb,amsfonts}
\usepackage{multirow}
\usepackage{array}
\usepackage{algorithmic}
\usepackage{graphicx}
\usepackage{textcomp}
\usepackage{xcolor}
\def\BibTeX{{\rm B\kern-.05em{\sc i\kern-.025em b}\kern-.08em
    T\kern-.1667em\lower.7ex\hbox{E}\kern-.125emX}}
\begin{document}

\title{SpO$_2$ Predictor-Guided Stage-Wise Time-Frequency Reconstruction of Low-Quality Dual-Wavelength PPG for Oxygen Saturation Estimation
\\
}


\author{
\IEEEauthorblockN{
Zequan Liang$^{1}$,
Elahe Hosseini$^{2}$,
Ning Miao$^{2}$,
Mahdi Pirayesh Shirazi Nejad$^{2}$,
Wei Shao$^{1}$,\\
Ehsan Kourkchi$^{2}$,
Setareh Rafatirad$^{1}$,
Houman Homayoun$^{2}$
} \\
\IEEEauthorblockA{
$^{1}$Department of Computer Science, University of California, Davis, Davis, CA, U.S.A.\\  
$^{2}$Department of Electrical and Computer Engineering, University of California, Davis, Davis, CA, U.S.A.\\ 
Email: \{zqliang, ehosseini, nmiao, pirayesh, wayshao, ekay, srafatirad, hhomayoun\}@ucdavis.edu}} 

\maketitle

\begin{abstract}
Continuous oxygen saturation (SpO$_2$) estimation from wearable photoplethysmography (PPG) is important for long-term health monitoring, but low-quality red and infrared PPG segments can distort waveform morphology and degrade SpO$_2$ prediction accuracy. Existing PPG denoising and reconstruction methods usually optimize waveform fidelity or heart rate characteristics, while time-domain waveform loss on PPG signals alone insufficiently preserves frequency structure and SpO$_2$-relevant information. This paper proposes a SpO$_2$ predictor-guided stage-wise time-frequency reconstruction framework for low-quality dual-wavelength PPG signals. The proposed method first selects high-quality PPG segments to pretrain a SpO$_2$ predictor. A masked reconstruction model is then trained to recover randomly masked PPG regions using a joint reconstruction objective that combines time-domain waveform loss with frequency-domain loss computed from the short-time Fourier transform (STFT). To make the reconstruction task physiologically relevant, the pretrained SpO$_2$ predictor is incorporated as an additional constraint, encouraging the reconstructed PPG to preserve SpO$_2$ information rather than only minimizing waveform reconstruction error. The SpO$_2$ predictor and PPG reconstructor model are optimized through four training stages. Experiments on the public OpenOximetry Repository and a private wearable PPG dataset show that the proposed approach achieves the lowest subject-level MAE, with 2.882\% on the public dataset and 2.359\% on the private dataset.

\end{abstract}


\section{Introduction}

Oxygen saturation (SpO$_2$) is an important physiological indicator for monitoring respiratory and cardiovascular status \cite{evans2001vital}. Conventional pulse oximetry estimates SpO$_2$ from dual-wavelength photoplethysmography (PPG) signals, typically red and infrared PPG, by exploiting their wavelength-dependent absorption differences and computing the ratio between pulsatile and baseline components \cite{kumar2021pulse}. However, wearable PPG acquisition is often affected by motion artifacts, unstable skin contact, and variations in optical coupling \cite{11337749}. These factors can distort the morphology and frequency characteristics of red and infrared PPG signals, thereby reducing the reliability of downstream SpO$_2$ estimation.

Existing methods for handling low-quality PPG signals commonly rely on adaptive filtering, segment rejection, or signal reconstruction. Adaptive filtering methods can suppress motion-related noise by using accelerometer signals as reference inputs to estimate and remove motion artifacts, but their performance depends on the availability and reliability of auxiliary motion measurements \cite{han2007development}. Segment rejection improves input reliability by discarding low-quality data based on signal quality assessment (SQA), but it can be overly conservative and may remove segments that still contain useful physiological information \cite{mohagheghian2022optimized}. Machine learning reconstruction methods provide a more flexible way to restore degraded PPG signals, but many existing objectives mainly focus on time-domain waveform fidelity or the preservation of heart-rate-related characteristics \cite{jain2024self,wang2022ppg}. Moreover, supervised reconstruction methods often require clean reference signals, which are difficult to obtain in practice when the acquired source signals are already degraded \cite{gautam2024autoencoder}.

\begin{figure}[t]
\centering
\includegraphics[width=1\linewidth]{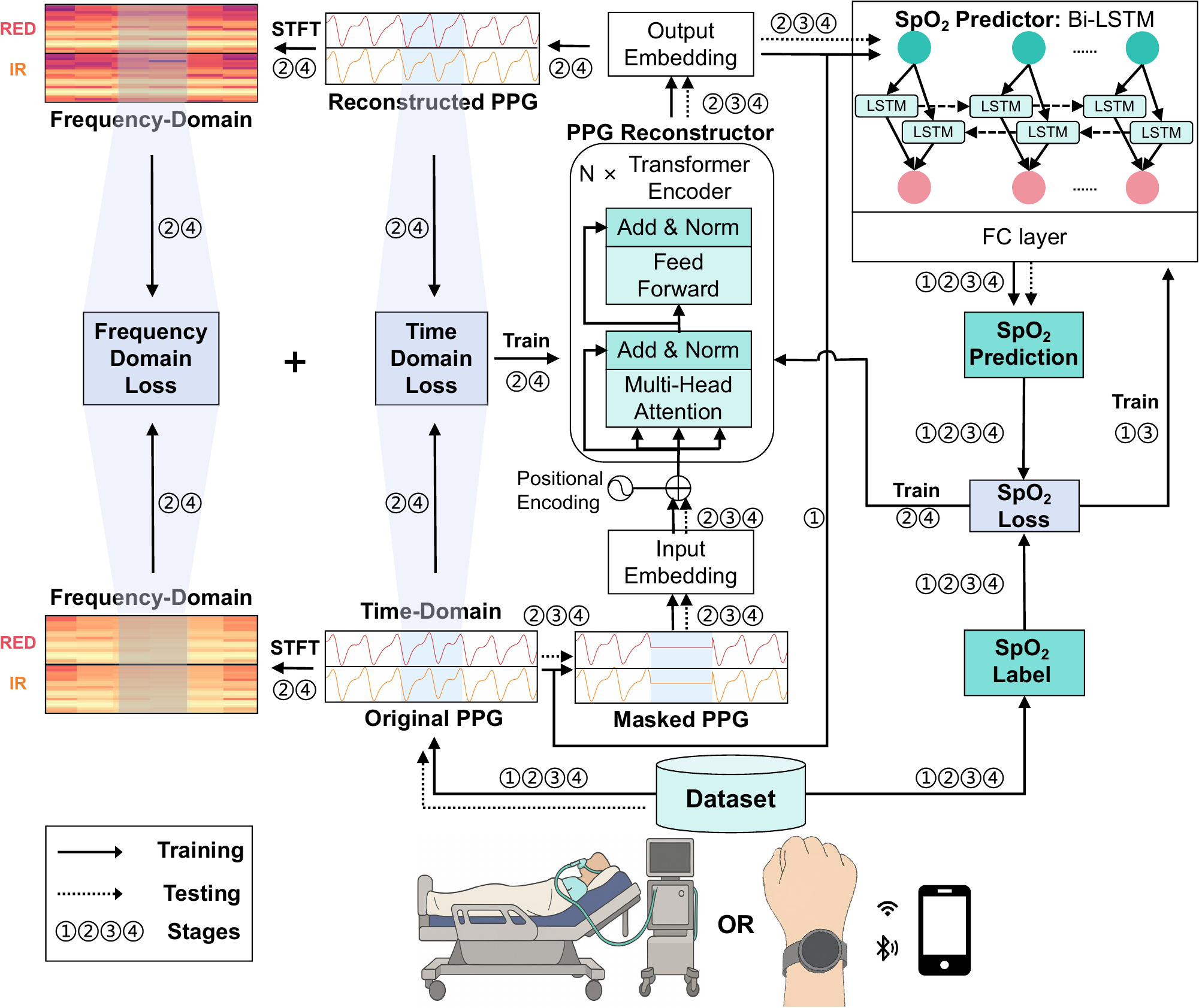}
\caption{Overview of the proposed SpO$_2$ predictor-guided stage-wise time-frequency PPG reconstruction framework. }
\label{fig:spo2_framework_flowchart}
\end{figure}

For SpO$_2$ estimation based on dual-wavelength PPG signals, the downstream target depends not only on the quality of each individual waveform, but also on whether the reconstructed red and infrared signals preserve SpO$_2$-relevant morphology, amplitude relationships, and spectral structure. Therefore, reconstruction should not be treated only as a generic denoising or signal restoration problem. Instead, the reconstruction objective should be aligned with the downstream physiological estimation task \cite{badiola2025real}. This motivates an SpO$_2$ predictor-guided reconstruction strategy, in which a pretrained SpO$_2$ estimation model provides an additional physiological constraint during PPG reconstruction.

In this work, we propose an SpO$_2$ predictor-guided stage-wise time-frequency reconstruction framework for low-quality dual-wavelength PPG signals to improve SpO$_2$ estimation. As illustrated in Fig.~\ref{fig:spo2_framework_flowchart}, the proposed framework first selects high-quality PPG segments to pretrain an SpO$_2$ predictor. Then, inspired by the masked autoencoding strategy~\cite{he2022masked}, we train a masked PPG reconstruction model to recover randomly masked temporal regions from the remaining signal context. The reconstructor is optimized using a joint reconstruction objective that combines a time-domain waveform loss and a frequency-domain loss computed from the short-time Fourier transform (STFT)~\cite{allen2005unified}. To make the reconstructed signals physiologically relevant, the pretrained SpO$_2$ predictor is incorporated as an additional constraint, encouraging the reconstruction model to preserve SpO$_2$-related information rather than merely minimizing waveform-level reconstruction error. The SpO$_2$ predictor and PPG reconstructor  are optimized through four training stages to progressively improve both SpO$_2$ estimation and PPG reconstruction.

The main contributions of this work are summarized as follows:

\begin{itemize}
\item We propose an SpO$_2$ predictor-guided reconstruction framework for low-quality dual-wavelength PPG signals, where a pretrained SpO$_2$ predictor is used as a physiological constraint to guide PPG reconstruction and improve oxygen saturation estimation.

\item We introduce a joint time-frequency reconstruction objective that combines time-domain waveform loss with frequency-domain loss computed from STFT, enabling the reconstruction model to preserve both PPG morphology and frequency structure.

\item We design a stage-wise optimization strategy that alternates between SpO$_2$ predictor training and masked PPG reconstruction model training, enabling joint improvement of downstream SpO$_2$ estimation and PPG reconstruction. Experiments on the public OpenOximetry Repository \cite{fong2025open} and a private wearable dataset \cite{11337411} show that the proposed approach achieves the best downstream SpO$_2$ estimation performance.
\end{itemize}

\section{Proposed Method}
As shown in Fig.~\ref{fig:spo2_framework_flowchart}, the proposed SpO$_2$ predictor-guided stage-wise time-frequency PPG reconstruction framework consists of four training stages:
\begin{itemize}
    \item \textbf{Stage 1: SpO$_2$ predictor pretraining.} 
    The Bi-LSTM with attention SpO$_2$ predictor \cite{11337411} is pretrained using the original high-quality PPG segments and the corresponding SpO$_2$ labels.

    \item \textbf{Stage 2: Predictor-guided PPG reconstructor
    training.} 
    The transformer-based PPG reconstructor \cite{vaswani2017attention} is trained to reconstruct the original high-quality PPG segments from inputs with randomly masked regions. 
    The reconstructor is optimized using both time-domain and frequency-domain reconstruction losses. 
    In addition, the pretrained SpO$_2$ predictor is fixed and connected after the PPG reconstructor to provide an SpO$_2$ prediction loss.

    \item \textbf{Stage 3: SpO$_2$ predictor refinement.} 
    The SpO$_2$ predictor is further trained using reconstructed PPG representations generated from all PPG segments by the fixed reconstructor.

    \item \textbf{Stage 4: PPG reconstructor refinement.} 
    The PPG reconstructor is further refined on high-quality PPG segments using the same joint objective as in Stage~2.
\end{itemize}

\subsection{PPG Segment Preprocessing}
\begin{figure}[t]
\centering
\includegraphics[width=0.8\linewidth]{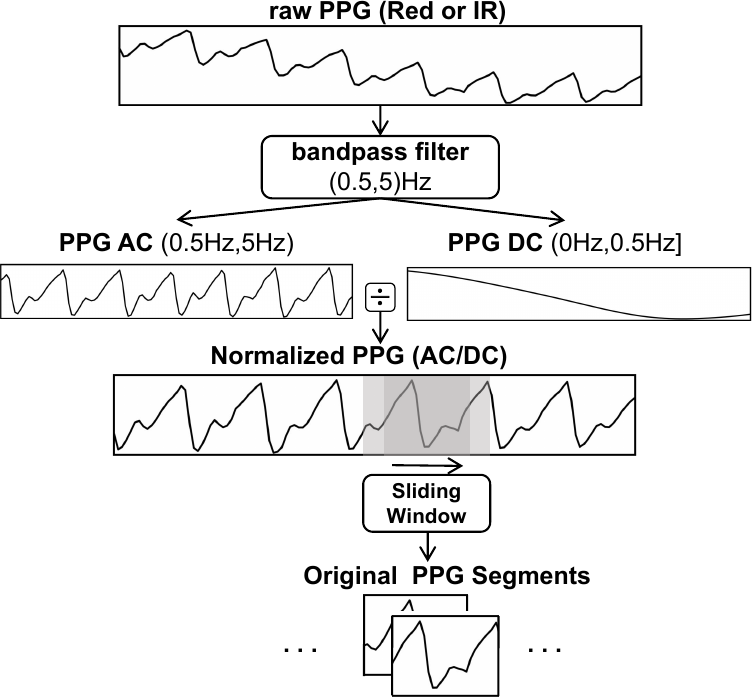}
\caption{PPG segment preprocessing. }
\label{fig:ppg_preprocess}
\end{figure}

To avoid subject leakage, the training set is constructed by splitting the data at the subject level.
For each subject, as shown in Fig.~\ref{fig:ppg_preprocess}, the raw red and infrared PPG signals are decomposed into alternating current (AC) components and slowly varying direct current (DC) components.
The AC component is extracted using a band-pass filter within 0.5--5 Hz, while the DC component is obtained from the low-frequency baseline component below 0.5 Hz.
The normalized PPG signal is then computed as the ratio between the AC and DC components (AC/DC) and segmented into 10-s windows with a 1-s sliding step to generate the original PPG segments for subsequent model inputs \cite{11337411}.

\subsection{Stage 1: SpO$_2$ Predictor Pretraining}

The SpO$_2$ predictor is pretrained to provide reliable SpO$_2$ estimation from high-quality dual-wavelength PPG signals. 
High-quality normalized AC/DC PPG segments  are selected using a template-matching PPG signal quality assessment (SQA) method \cite{orphanidou2017quality} implemented in NeuroKit2 \cite{makowski2021neurokit2}, which assigns a quality score between 0 and 1 to each sample.
Specifically, a segment is retained only if all 1-s average quality scores of the red-wavelength signal are not lower than 0.6.

Let $\mathbf{x}_i \in \mathbb{R}^{T \times C}$ denote the $i$-th original dual-channel PPG segment, where $T$ is the number of time samples and $C=2$ denotes the red and infrared channels.
Let $y_i$ denote the reference SpO$_2$ value aligned with the segment, and let $P(\cdot)$ denote the SpO$_2$ predictor.

The predictor is optimized using the mean square error (MSE) loss:
\begin{equation}
\mathcal{L}_{\mathrm{SpO}_2}
=
\frac{1}{N}
\sum_{i=1}^{N}
\left(
P(\mathbf{x}_i) - y_i
\right)^2
\end{equation}
where $N$ denotes the number of training segments in a mini-batch.

\subsection{Stage 2: Predictor-Guided PPG Reconstructor Training}

In the second stage, only the high-quality PPG segments selected in Stage~1 are used to train the PPG reconstructor. 
For each dual-channel PPG segment, a contiguous temporal region with a random duration from 1~s to 5~s is selected and set to zero, producing a masked input segment.
The masked PPG segment is first mapped to an input embedding through a linear projection, and positional encoding is then added before being fed into the PPG reconstructor, which consists of four stacked Transformer encoder layers, each including multi-head self-attention, feed-forward layers, residual connections, and normalization layers.
The output embeddings are projected back to the dual-channel PPG space to obtain the reconstructed PPG segment.

Given the masked input segment $\mathbf{x}_i^{\mathrm{mask}} \in \mathbb{R}^{T \times C}$, the PPG reconstructor $R(\cdot)$ generates the reconstructed PPG segment as $\hat{\mathbf{x}}_i = R(\mathbf{x}_i^{\mathrm{mask}})$, 
where $\hat{\mathbf{x}}_i \in \mathbb{R}^{T \times C}$ denotes the reconstructed dual-channel PPG segment.

\subsubsection{Time-Domain Reconstruction Loss}
To encourage accurate reconstruction of the masked temporal region, the reconstructed segment $\hat{\mathbf{x}}_i$ is constrained to be close to the original high-quality PPG segment $\mathbf{x}_i$ in the time domain. The time-domain reconstruction loss is computed over the masked samples as:
\begin{equation}
\mathcal{L}_{\mathrm{time}}
=
\frac{1}{N}
\sum_{i=1}^{N}
\frac{1}{|\Omega_i|C}
\sum_{t \in \Omega_i}
\sum_{c=1}^{C}
\left(
\hat{x}_{i,t,c} - x_{i,t,c}
\right)^2
\end{equation}
where $\Omega_i$ denotes the set of masked time samples in the $i$-th segment, $|\Omega_i|$ is the number of masked time samples, $t$ denotes the PPG sample index, and $c$ denotes the PPG channel index.

\subsubsection{Frequency-Domain Reconstruction Loss}
To further preserve the time-frequency characteristics of PPG signals, the original and reconstructed segments, $\mathbf{x}_i$ and $\hat{\mathbf{x}}_i$, are transformed into the frequency domain using the short-time Fourier transform (STFT), yielding $\mathbf{S}_i, \hat{\mathbf{S}}_i \in \mathbb{R}^{T_{\mathrm{f}} \times F \times D}$, where $T_{\mathrm{f}}$ is the number of STFT time frames, $F$ is the number of frequency bins, and $D=4$ corresponds to the real and imaginary components of the red and infrared PPG channels.
To encourage consistency between $\mathbf{S}_i$ and $\hat{\mathbf{S}}_i$ over the masked region, the frequency-domain reconstruction loss is defined as:
\begin{equation}
\mathcal{L}_{\mathrm{freq}}
=
\frac{1}{N}
\sum_{i=1}^{N}
\frac{1}{|\Gamma_i|FD}
\sum_{\tau \in \Gamma_i}
\sum_{f=1}^{F}
\sum_{d=1}^{D}
\left(
\hat{S}_{i,\tau,f,d}
-
S_{i,\tau,f,d}
\right)^2
\end{equation}
where $\Gamma_i$ denotes the set of STFT time frames whose analysis windows overlap with the temporal masked region $\Omega_i$, $|\Gamma_i|$ is the number of these STFT time frames, $\tau$ denotes the STFT time-frame index, $f$ denotes the frequency-bin index, and $d$ indexes the real and imaginary components of the two PPG channels. The STFT parameters were set to an FFT size of 200, a hop length of 20, and a window length of 200.

\subsubsection{Predictor-Guided SpO$_2$ Loss}
After the reconstructed PPG segment $\hat{\mathbf{x}}_i = R(\mathbf{x}_i^{\mathrm{mask}})$ is generated, the recovered masked region is merged with the original unmasked region before being fed into the frozen pretrained SpO$_2$ predictor $P(\cdot)$. 
Specifically, the final merged reconstructed PPG segment $\bar{\mathbf{x}}_i$ is defined as
\begin{equation}
\bar{x}_{i,t,c} =
\begin{cases}
\hat{x}_{i,t,c}, & t \in \Omega_i, \\
x_{i,t,c}, & t \notin \Omega_i,
\end{cases}
\end{equation}

The predictor-guided SpO$_2$ loss is then defined as:
\begin{equation}
\mathcal{L}_{\mathrm{SpO}_2}^{\mathrm{guide}}
=
\frac{1}{N}
\sum_{i=1}^{N}
\left(
P(\bar{\mathbf{x}}_i) - y_i
\right)^2 .
\end{equation}
During the second stage, this loss updates only the PPG reconstructor $R(\cdot)$, while the pretrained SpO$_2$ predictor $P(\cdot)$ remains frozen.

\subsubsection{Overall Reconstruction Loss}
The overall training objective for Stage~2 is defined as:
\begin{equation}
\mathcal{L}_{\mathrm{recon}}
=
\mathcal{L}_{\mathrm{time}}
+
\lambda_{\mathrm{freq}}
\mathcal{L}_{\mathrm{freq}}
+
\lambda_{\mathrm{SpO}_2}
\mathcal{L}_{\mathrm{SpO}_2}^{\mathrm{guide}}
\label{eq}
\end{equation}
where $\lambda_{\mathrm{freq}}$ and $\lambda_{\mathrm{SpO}_2}$ control the contributions of the frequency-domain loss and the predictor-guided SpO$_2$ loss, respectively. In this work, $\lambda_{\mathrm{freq}}$ and $\lambda_{\mathrm{SpO}_2}$ were empirically selected by monitoring the magnitudes of the individual loss terms during the first training epoch, so that the weighted loss components remained on the same order of magnitude and their contributions to the overall objective were reasonably balanced.

This objective encourages the reconstructor to recover PPG signals that are close to the original waveform while preserving time-frequency characteristics and physiological information relevant to SpO$_2$ estimation.

\subsection{Stage 3: SpO$_2$ Predictor Refinement}

In the third stage, the pretrained PPG reconstructor is frozen, while the SpO$_2$ predictor is further optimized using the entire training set, including both high-quality and low-quality PPG segments.
For each input segment $\mathbf{x}_i$, the temporal region with the lowest $k=3$~s averaged quality score is selected as the masked region, producing the masked input segment $\mathbf{x}_i^{\mathrm{mask}}$.
The masked segment is then processed by the frozen PPG reconstructor $R(\cdot)$ to generate $\hat{\mathbf{x}}_i = R(\mathbf{x}_i^{\mathrm{mask}})$. Following the same merging operation defined in Stage~2, the merged reconstructed segment is $\bar{\mathbf{x}}_i$.

The SpO$_2$ predictor is updated by minimizing the SpO$_2$ estimation error from the reconstructed PPG segment:
\begin{equation}
\mathcal{L}_{\mathrm{SpO}_2}^{\mathrm{train}}
=
\frac{1}{N}
\sum_{i=1}^{N}
\left(
P(\bar{\mathbf{x}}_i) - y_i
\right)^2
\end{equation}
During this stage, only the SpO$_2$ predictor $P(\cdot)$ is updated, while the PPG reconstructor $R(\cdot)$ remains frozen.

This stage adapts the SpO$_2$ predictor to reconstructed PPG representations under both high-quality and low-quality signal conditions, improving its compatibility with the reconstructed PPG input to predict SpO$_2$.

\subsection{Stage 4: PPG Reconstructor Refinement}

In the fourth stage, the refined SpO$_2$ predictor is frozen, and the PPG reconstructor is further optimized on high-quality segments using the same objective as in Eq.~\eqref{eq}. During this stage, the frozen predictor provides physiological guidance for reconstructor refinement, encouraging the reconstructor to recover PPG signal components that are informative for downstream SpO$_2$ estimation.

\subsection{Testing Procedure}

During testing, the test subjects are processed independently and are not involved in any training stage.
The red and infrared PPG signals are segmented into 10-s windows with a 1-s sliding step.
For each test segment, regardless of its signal quality, a $k=3$~s temporal region with the lowest averaged quality score on the red-channel PPG signal is masked by setting its values to zero, producing the masked input segment $\mathbf{x}_i^{\mathrm{mask}}$.

The final PPG reconstructor first generates 
$\hat{\mathbf{x}}_i = R(\mathbf{x}_i^{\mathrm{mask}})$. 
The recovered masked region is then merged with the original unmasked region to obtain $\bar{\mathbf{x}}_i$.
The final SpO$_2$ predictor estimates the SpO$_2$ value from the merged reconstructed PPG signal:
\begin{equation}
\hat{y}_i = P(\bar{\mathbf{x}}_i).
\end{equation}
\section{Experiment Results}
\subsection{Dataset}

We evaluate the proposed framework for improving SpO$_2$ estimation from dual-wavelength PPG signals using two datasets: the public OpenOximetry dataset \cite{fong2025open} and a private wearable dataset collected using the We-Be band \cite{DBLP:conf/ndss/ShaoLZFMKRHF26
}.

The OpenOximetry dataset is a publicly accessible pulse oximetry repository designed to support the evaluation of pulse oximeter performance under controlled laboratory and clinical settings.
In this work, we use red and infrared PPG recordings from 191 valid subjects in the OpenOximetry dataset.
The red and infrared PPG signals were originally recorded at a sampling rate of 86~Hz, and the corresponding reference SpO$_2$ values were obtained by averaging readings from multiple pulse oximeters.
For consistency with the private wearable dataset, the OpenOximetry PPG signals are upsampled from 86~Hz to 100~Hz before segmentation.
The dual-wavelength PPG signals are then segmented into 10-s windows with a 1-s sliding step, and each segment is paired with the reference SpO$_2$ value aligned to the corresponding timestamp.
To avoid subject leakage, the training, validation, and testing sets are split at the subject level, ensuring that segments from the same subject do not appear in different subsets.
The size of the OpenOximetry dataset is summarized in Table~\ref{tab:openox_dataset}. In addition, the testing segment distribution across different PPG signal quality score ranges is shown in Fig.~\ref{fig:SQA_bins}.

\begin{table}[t]
\caption{Subject-level Split of the OpenOximetry Dataset}
\begin{center}
\begin{tabular}{|c|c|c|}
\hline
\textbf{Subject-level Split} & \textbf{Number of Subjects} & \textbf{Number of Segments} \\
\hline
Train & 124 & 261,820 \\
\hline
Validation & 32 & 68,777 \\
\hline
Test & 35 & 74,453 \\
\hline
Total & 191 & 405,050 \\
\hline
\end{tabular}
\label{tab:openox_dataset}
\end{center}
\end{table}

The private wearable dataset was collected using the wrist-based We-Be band for red and infrared PPG acquisition at 100~Hz, with reference SpO$_2$ values measured by a Masimo Rad-G fingertip pulse oximeter.
A total of 9 participants completed three data-collection sessions, and each session included three breath-holding trials to induce transient SpO$_2$ fluctuations.
The collected PPG signals were segmented into 12,472 10-s windows, each aligned with the corresponding SpO$_2$ reading.
Due to the limited number of subjects, the private wearable dataset was evaluated using a leave-one-subject-out (LOSO) strategy, with the average performance reported across all held-out subjects. In each LOSO fold, models pretrained on the public OpenOximetry dataset were fine-tuned only on the training subjects from the private wearable dataset and then evaluated on the held-out subject.

\subsection{Ablation Study}

We conduct an ablation study to evaluate the contribution of the three main components in the proposed PPG reconstruction framework: the time-domain reconstruction loss $\mathcal{L}_{\mathrm{time}}$, the frequency-domain reconstruction loss $\mathcal{L}_{\mathrm{freq}}$, and the predictor-guided SpO$_2$ loss $\mathcal{L}_{\mathrm{SpO}_2}^{\mathrm{guide}}$.
Each component is removed individually, and SpO$_2$ estimation performance is evaluated using mean absolute error (MAE) and root mean square error (RMSE) at both the segment and subject levels under the $k=3$~s masking setting. Segment-level metrics are computed over all test segments. Subject-level metrics are first computed separately for each test subject using all segments from that subject and are then averaged across test subjects.

\begin{table}[t]
\caption{SpO$_2$ estimation performance of different ablation variants on the OpenOximetry Dataset.}
\begin{center}
\begin{tabular}{|c|c|c|c|c|c|c|}
\hline
\multirow{2}{*}{\boldmath{$\mathcal{L}_{\mathrm{time}}$} }
& \multirow{2}{*}{\boldmath{$\mathcal{L}_{\mathrm{freq}}$}} 
& \multirow{2}{*}{\boldmath{$\mathcal{L}_{\mathrm{SpO}_2}^{\mathrm{guide}}$}} 
& \multicolumn{2}{c|}{\textbf{Subject}} 
& \multicolumn{2}{c|}{\textbf{Segment}} \\
\cline{4-7}
& & & \textbf{\textit{MAE}} & \textbf{\textit{RMSE}} & \textbf{\textit{MAE}} & \textbf{\textit{RMSE}} \\
\hline
 & $\checkmark$ & $\checkmark$ & 2.935 & 4.320 & 3.026 & \textbf{4.911} \\
\hline
$\checkmark$ &  & $\checkmark$ & 2.948 & 4.438 & 3.014 & 4.919 \\
\hline
$\checkmark$ & $\checkmark$ &  & 3.071 & 4.483 & 3.162 & 4.993 \\
\hline
$\checkmark$ & $\checkmark$ & $\checkmark$ & \textbf{2.882} & \textbf{4.297} & \textbf{2.974} & 4.913 \\
\hline
\end{tabular}
\label{tab:ablation_study_public}
\end{center}
\end{table}

As shown in Table~\ref{tab:ablation_study_public}, incorporating all three components achieves the best SpO$_2$ estimation performance on the OpenOximetry dataset across most evaluation metrics, except for segment-level RMSE.
Notably, the variant without the predictor-guided SpO$_2$ loss performs worse than the variants without either the time-domain or frequency-domain reconstruction loss across all evaluation metrics.
This indicates that the physiological guidance provided by the SpO$_2$ predictor is the most critical component for downstream SpO$_2$ estimation.
Removing either the time-domain or frequency-domain reconstruction loss also degrades performance, suggesting that accurate waveform reconstruction in the time domain and consistent spectral representation in the frequency domain are both important for PPG reconstruction and ultimately benefit downstream SpO$_2$ estimation.
Overall, the ablation results demonstrate that the proposed joint objective provides the most consistent improvement for SpO$_2$ estimation at both the subject and segment levels.

\begin{table}[t]
\caption{Reconstruction MAE of different ablation variants in the time and frequency domains on the OpenOximetry dataset.}
\setlength{\tabcolsep}{4.5pt}
\begin{center}
\begin{tabular}{|c|c|c|c|c|c|c|}
\hline
\multirow{2}{*}{\boldmath{$\mathcal{L}_{\mathrm{time}}$}}
& \multirow{2}{*}{\boldmath{$\mathcal{L}_{\mathrm{freq}}$}}
& \multirow{2}{*}{\boldmath{$\mathcal{L}_{\mathrm{SpO}_2}^{\mathrm{guide}}$}}
& \multicolumn{2}{c|}{\textbf{Time}}
& \multicolumn{2}{c|}{\textbf{Frequency}} \\
\cline{4-7}
& & & \textbf{\textit{Red}} & \textbf{\textit{IR}} & \textbf{\textit{Red}} & \textbf{\textit{IR}} \\
\hline

 & $\checkmark$ & $\checkmark$
& 0.00268 & 0.00329
& 0.00722 & 0.00876 \\
\hline

$\checkmark$ &  & $\checkmark$
& 0.00246 & 0.00308
& 0.00933 & 0.01176 \\
\hline

$\checkmark$ & $\checkmark$ & 
& \textbf{0.00218} & \textbf{0.00271}
& 0.00451 & 0.00550 \\
\hline

$\checkmark$ & $\checkmark$ & $\checkmark$
& 0.00223 & 0.00274
& \textbf{0.00446} & \textbf{0.00531} \\
\hline

\end{tabular}
\label{tab:reconstruction_mae_private}
\end{center}
\end{table}

On the OpenOximetry dataset, we also report the reconstruction MAE between the reconstructed high-quality PPG segments and the original PPG segments in both the time and frequency domains for the red and IR channels, as shown in Table~\ref{tab:reconstruction_mae_private}. 
Removing the time-domain loss increases the time-domain reconstruction error, while removing the frequency-domain loss increases the frequency-domain reconstruction error, indicating that the two reconstruction objectives provide complementary supervision. 
The variant using both $\mathcal{L}_{\mathrm{time}}$ and $\mathcal{L}_{\mathrm{freq}}$ achieves the lowest time-domain MAE, whereas the full framework achieves the lowest frequency-domain MAE for both red and IR channels. 
These results suggest that the predictor-guided SpO$_2$ loss improves downstream SpO$_2$ estimation while introducing only a small and acceptable increase in time-domain reconstruction error.

\begin{table}[t]
\caption{SpO$_2$ estimation performance of different ablation variants on the private wearable dataset.}
\begin{center}
\begin{tabular}{|c|c|c|c|c|c|c|}
\hline
\multirow{2}{*}{\boldmath{$\mathcal{L}_{\mathrm{time}}$} }
& \multirow{2}{*}{\boldmath{$\mathcal{L}_{\mathrm{freq}}$}} 
& \multirow{2}{*}{\boldmath{$\mathcal{L}_{\mathrm{SpO}_2}^{\mathrm{guide}}$}} 
& \multicolumn{2}{c|}{\textbf{Subject}} 
& \multicolumn{2}{c|}{\textbf{Segment}} \\
\cline{4-7}
& & & \textbf{\textit{MAE}} & \textbf{\textit{RMSE}} 
& \textbf{\textit{MAE}} & \textbf{\textit{RMSE}} \\
\hline
 & $\checkmark$ & $\checkmark$ & 2.401 & 3.112 & 2.481 & 3.463 \\
\hline
$\checkmark$ &  & $\checkmark$ & 2.478 & 3.101 & 2.552 & 3.428 \\
\hline
$\checkmark$ & $\checkmark$ &  & 2.522 & 3.179 & 2.665 & 3.600 \\
\hline
$\checkmark$ & $\checkmark$ & $\checkmark$ 
& \textbf{2.359} & \textbf{2.973} & \textbf{2.468} & \textbf{3.338} \\
\hline
\end{tabular}
\label{tab:ablation_study_private}
\end{center}
\end{table}

Regarding the SpO$_2$ estimation results in Table~\ref{tab:ablation_study_private}, the full framework incorporating all three components achieves the best performance on the private wearable dataset. Similar to the public OpenOximetry dataset, removing the predictor-guided SpO$_2$ loss leads to the largest performance degradation, while removing either the time-domain or frequency-domain reconstruction loss also reduces performance. These results are consistent with the OpenOximetry ablation study and further validate the effectiveness of the proposed joint objective under wearable PPG conditions.

\subsection{Comparison with Related Methods}

We compare the proposed method with related methods on the public OpenOximetry dataset. 
For all compared methods, the input PPG signals are preprocessed to 100~Hz and segmented into 10-s windows. 
To ensure a common evaluation setting across all methods, we report subject-level MAE and RMSE.

\begin{itemize}
    \item \textbf{Calibration}~\cite{guo2015reflective}: 
    A conventional calibration method that computes the ratio-of-ratios $R$ from the red and infrared PPG signals and estimates SpO$_2$ using a quadratic calibration function.

    \item \textbf{NormWear}~\cite{luo2026toward}: 
    A pretrained PPG foundation model is used to extract PPG embeddings, followed by a regression model for SpO$_2$ estimation.

    \item \textbf{Baseline}~\cite{11337411}: 
    A Bi-LSTM with attention is trained as the SpO$_2$ predictor and directly estimates SpO$_2$ from the original PPG segments without PPG reconstruction.

    \item \textbf{Proposed}: 
    The proposed SpO$_2$ predictor-guided stage-wise time-frequency PPG reconstruction framework for SpO$_2$ estimation.
\end{itemize}

\begin{table}[b]
\caption{Subject-level SpO$_2$ estimation performance comparison with related methods on the OpenOximetry dataset.}
\label{tab:comparison}
\begin{center}
\begin{tabular}{|c|c|c|}
\hline
\multirow{2}{*}{\textbf{Method}}
& \multicolumn{2}{c|}{\textbf{Subject}} \\
\cline{2-3}
& \textbf{\textit{MAE}} 
& \textbf{\textit{RMSE}} \\
\hline
Calibration~\cite{guo2015reflective} & 3.460 & 4.847 \\
\hline
NormWear~\cite{luo2026toward} & 3.390 & 5.430 \\
\hline
Baseline~\cite{11337411} & 3.063 & 4.506 \\
\hline
Proposed & \textbf{2.882} & \textbf{4.297} \\
\hline
\end{tabular}
\end{center}
\end{table}

\begin{table}[b]
\caption{SpO$_2$ estimation performance of different training stages on the OpenOximetry dataset.}
\begin{center}
\begin{tabular}{|c|c|c|c|c|}
\hline
\multirow{2}{*}{\textbf{Stage}}
& \multicolumn{2}{c|}{\textbf{Subject}} 
& \multicolumn{2}{c|}{\textbf{Segment}} \\
\cline{2-5}
& \textbf{\textit{MAE}} 
& \textbf{\textit{RMSE}} 
& \textbf{\textit{MAE}} 
& \textbf{\textit{RMSE}} \\
\hline
Stage~2 & 3.370 & 5.140 & 3.392 & 5.289 \\
\hline
Stage~3 & 2.968 & 4.379 & 3.056 & 4.983 \\
\hline
Stage~4 & \textbf{2.882} & \textbf{4.297} & \textbf{2.974} & \textbf{4.913} \\
\hline
\end{tabular}
\label{tab:stage_performance_public}
\end{center}
\end{table}

As shown in Table~\ref{tab:comparison}, the proposed method achieves the lowest subject-level MAE and RMSE among all compared methods, showing that predictor-guided PPG reconstruction improves downstream SpO$_2$ estimation beyond direct prediction from the original PPG input.

\subsection{Stage-wise Analysis}
To demonstrate the necessity of multi-stage training, Table~\ref{tab:stage_performance_public} reports the SpO$_2$ estimation performance after different training stages. Stage~1 is excluded from this comparison because it only pretrains the SpO$_2$ predictor using high-quality PPG segments. The results show a consistent improvement from Stage~2 to Stage~4 at both the subject and segment levels. This indicates that the stage-wise strategy helps train the PPG reconstructor and SpO$_2$ predictor separately with specific objectives, while jointly improving the final SpO$_2$ estimation accuracy.

\subsection{Mask Duration Analysis}
\begin{figure}[t]
\centering
\includegraphics[width=\linewidth]{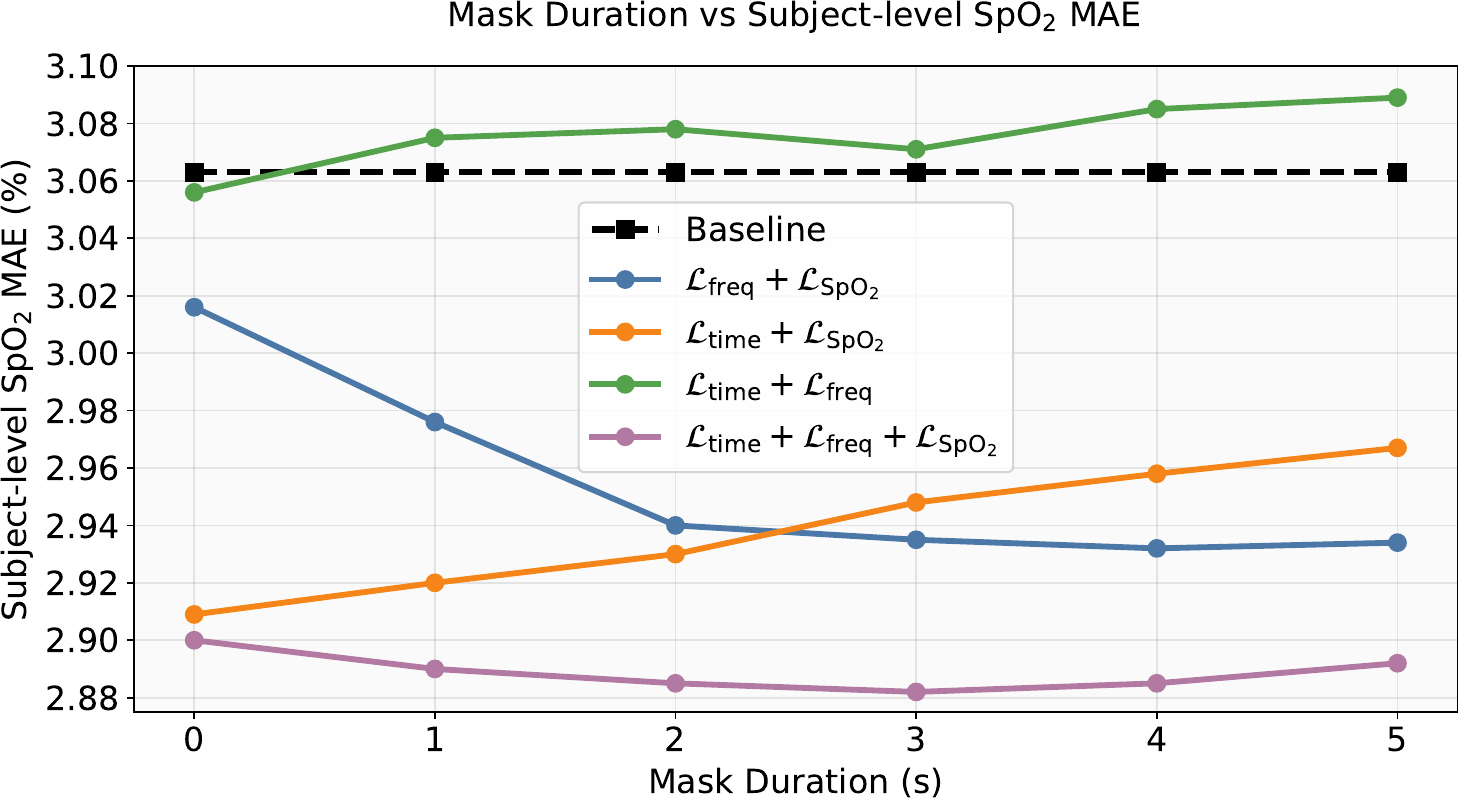}
\caption{Sensitivity analysis of different mask durations on subject-level SpO$_2$ MAE in the OpenOximetry dataset.}
\label{fig:mask_duration}
\end{figure}

During Stage~3 training and testing, a fixed $k$-second temporal region with the lowest averaged quality score is masked and reconstructed. 
Since the PPG reconstructor is pretrained by reconstructing randomly masked regions with durations from 1~s to 5~s within each 10-s segment, we evaluate the sensitivity of the framework to different mask durations, as shown in Fig.~\ref{fig:mask_duration}. 
We also include the 0-s masking setting, where no temporal region is reconstructed, and the final SpO$_2$ predictor is directly evaluated on the original PPG segment.

As shown in Fig.~\ref{fig:mask_duration}, the full framework achieves the best overall performance across different mask durations and obtains the lowest SpO$_2$ MAE when $k=3$ in the OpenOximetry dataset. 
Therefore, a mask duration of $k=3$~s is selected for the other evaluation. 
As the mask duration increases, the SpO$_2$ estimation errors of the variant without the time-domain loss and the full framework first decrease and then slightly increase, suggesting an empirical trade-off between the amount of reconstructed PPG signal and the preservation of SpO$_2$ information. 
In contrast, the errors of the other two variants generally increase with longer mask durations, indicating weaker robustness when either frequency-domain reconstruction supervision or predictor-guided SpO$_2$ supervision is removed.

When any component is removed, the SpO$_2$ estimation error increases across different masked regions. 
In particular, removing the predictor-guided SpO$_2$ loss leads to performance worse than the simple  SpO$_2$ predictor baseline, indicating that this component provides an important constraint for reconstruction toward the downstream SpO$_2$ estimation task. 

\begin{figure}[t]
\centering
\includegraphics[width=\linewidth]{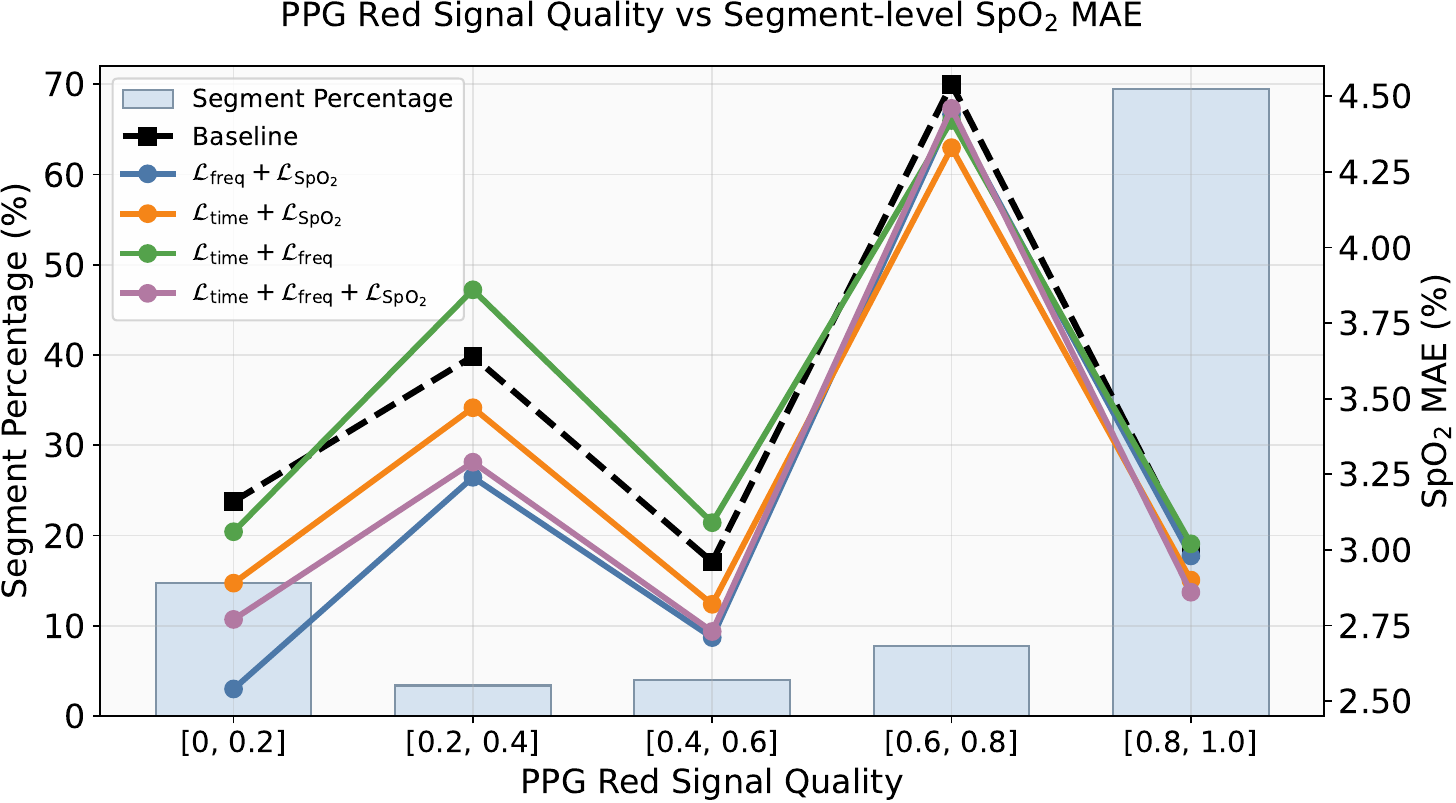}
\caption{Segment-level MAE and testing segment distribution across different PPG signal quality scores in the OpenOximetry dataset.}
\label{fig:SQA_bins}
\end{figure}

\begin{figure}[t]
\centering
\includegraphics[width=\linewidth]{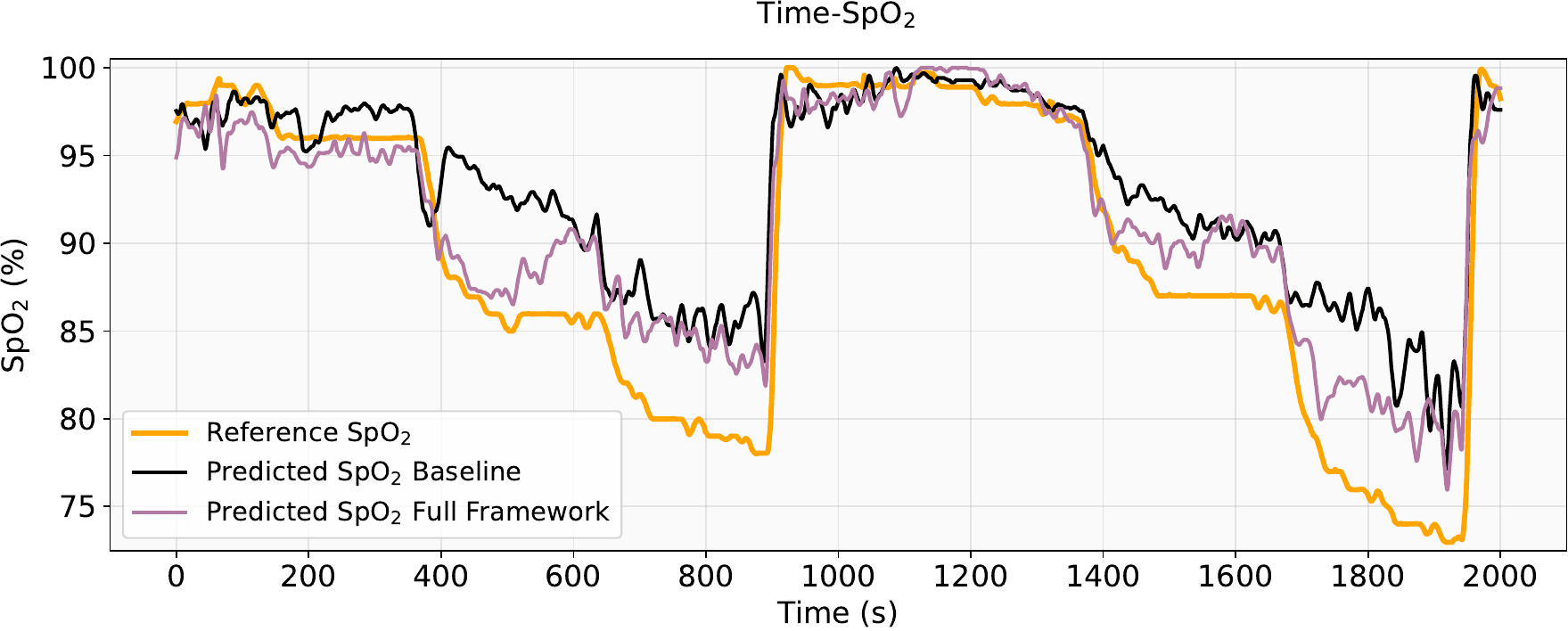}
\caption{Representative SpO$_2$ estimation case from the public OpenOximetry dataset. }
\label{fig:spo2_case_public}
\end{figure}

\subsection{Signal Quality Analysis}
We further analyze the segment-level SpO$_2$ MAE across different PPG red signal quality scores under the selected $k=3$~s masking setting. The segment distribution in each bin is also reported to show the data composition of the OpenOximetry dataset. As shown in Fig.~\ref{fig:SQA_bins}, most testing segments are distributed in the high-quality range. For lower-quality segments, especially those with quality scores below 0.6, the variants with the predictor-guided SpO$_2$ constraint achieve lower MAE than the baseline predictor. This suggests that the SpO$_2$-guided reconstruction objective helps preserve SpO$_2$ information during PPG reconstruction, improving downstream estimation accuracy on low-quality PPG segments.

\begin{figure}[t]
\centering
\includegraphics[width=\linewidth]{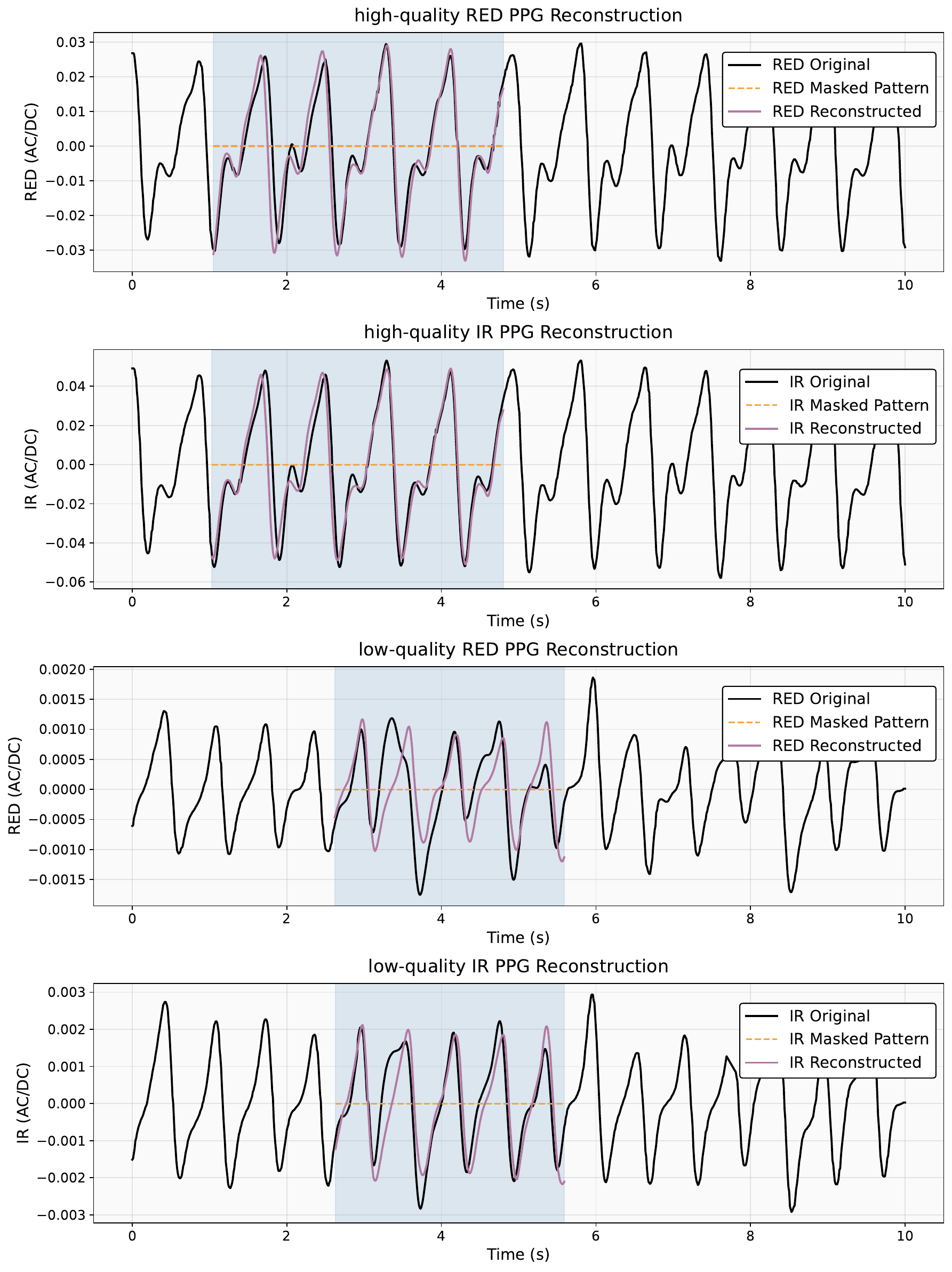}
\caption{Representative PPG reconstruction examples from the public OpenOximetry dataset.}
\label{fig:ppg_case_public}
\end{figure}

\subsection{Case Visualization}
We a visualize representative SpO$_2$ estimation case from the public OpenOximetry dataset in Fig.~\ref{fig:spo2_case_public}. Compared with the baseline predictor, the full proposed framework produces SpO$_2$ estimates that are closer to the reference SpO$_2$ labels at several time periods, showing that the proposed framework can improve the temporal agreement between predicted and reference SpO$_2$ values.


Regarding PPG reconstruction, we visualize representative masked regions from the public OpenOximetry dataset and compare the reconstructed signals with the corresponding original PPG signals. As shown in Fig.~\ref{fig:ppg_case_public}, the highlighted temporal regions in both the red and IR channels are masked. For the high-quality PPG segment, the reconstructed signals closely follow the waveform trend of the original signals within the masked interval. For the low-quality PPG segment, the original signals within the masked region exhibit visible waveform distortion, while the reconstructed signals show more regular morphology and smoother amplitude variation. It suggests a potential denoising effect, which may provide more reliable inputs for downstream SpO$_2$ estimation.

\section{Conclusion}

This paper proposed an SpO$_2$ predictor-guided stage-wise time-frequency reconstruction framework for low-quality dual-wavelength PPG signals. The proposed method trains a masked PPG reconstructor using a joint objective that combines time-domain loss, frequency-domain loss, and a predictor-guided SpO$_2$ loss. By incorporating a pretrained SpO$_2$ predictor as a physiological constraint, the reconstruction model is encouraged to preserve SpO$_2$ information for downstream oxygen saturation estimation. Experiments on the public OpenOximetry dataset and a private wearable PPG dataset showed that the proposed framework achieved the best overall SpO$_2$ estimation performance among the compared ablation variants. 

Future work will investigate adaptive masking and reconstruction strategies, where the mask duration and location are dynamically determined. We will also extend the predictor-guided reconstruction framework to other downstream tasks from PPG signals, such as heart-rate estimation and respiratory monitoring.
\bibliographystyle{ieeetr}
\bibliography{EMBC}

\end{document}